\documentclass[twocolumn,showpacs,preprintnumbers,amsmath,amssymb,prl]{revtex4}

\usepackage{graphicx}

\begin{document}

\title{Directional emission of stadium-shaped micro-lasers}

\author{M. Lebental$^{1,2}$, J. S. Lauret$^{1}$, J. Zyss$^1$\\C. Schmit$^2$, E. Bogomolny$^2$}
\email{lebental@lpqm.ens-cachan.fr} \affiliation{$^1$CNRS, \'Ecole
Normale Sup\'erieure de
Cachan,  UMR 8537, \\
Laboratoire de Photonique Quantique et Mol\'eculaire, 94235 Cachan, France \\
$^2$ CNRS, Universit\'e Paris Sud,  UMR 8626\\
Laboratoire de Physique Th\'eorique et Mod\`eles Statistiques, 91405
Orsay, France}

\date{\today}

\begin{abstract}
The far-field emission of two dimensional (2D) stadium-shaped
dielectric cavities is investigated. Micro-lasers with such shape
present a highly directional emission. We provide experimental
evidence of the dependance of the emission directionality on the
shape of the stadium, in good agreement with ray numerical
simulations. We develop a simple geometrical optics model which
permits to explain analytically main observed features.  Wave
numerical calculations confirm the results.
\end{abstract}


\maketitle

The field of quantum chaos has widely broadened over the last two
decades \cite{ecoles}. Generally, it relates the quantum behavior of
a broad diversity of systems with their classical features. In this
context, lossless billiards are models of great interest, providing
a broad variety of dynamical systems by changing the shape of the
boundary. Moreover they are accessible to experimental studies
\cite{microondes,Davidson}. In the same time, the influence of loss
or noise on open quantum systems is also being investigated
\cite{decoherence}, with emphasis on the behavior of open quantum
systems with chaotic classical dynamics \cite{cvitanovic,
boulanger,random}. Flat micro-lasers exhibiting different boundary
shapes are relevant examples of open billiards with a coherent
output coupling \cite{copains}.

\begin{figure}[ht!]
  \centering
    \includegraphics[width=.45\linewidth]{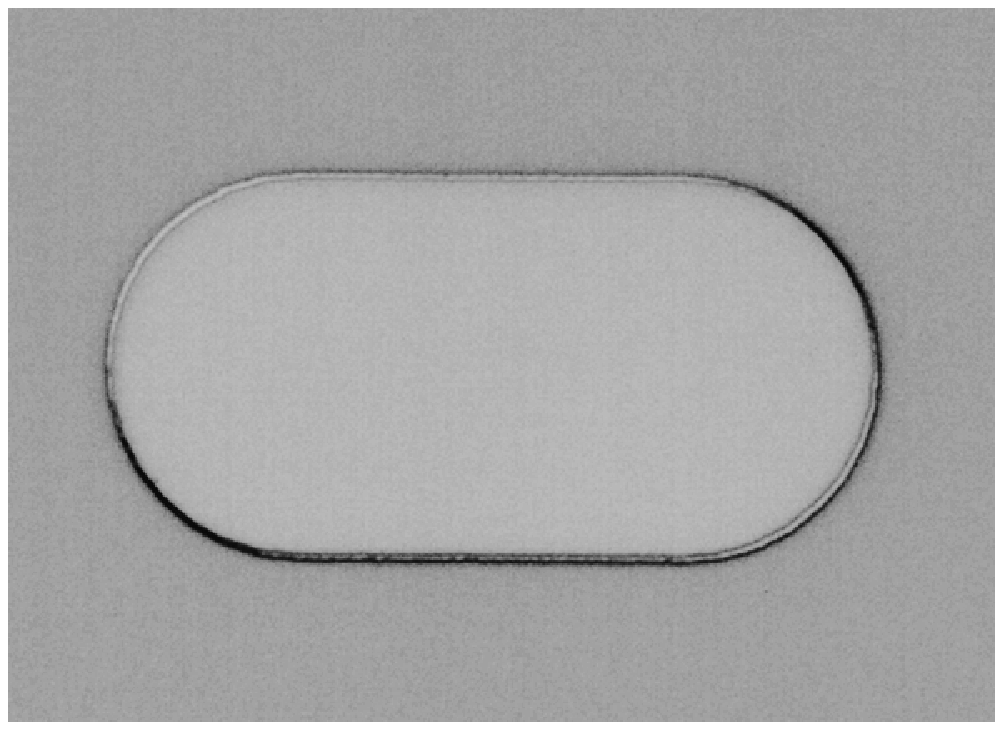}\hfill
    \includegraphics[width=.55\linewidth]{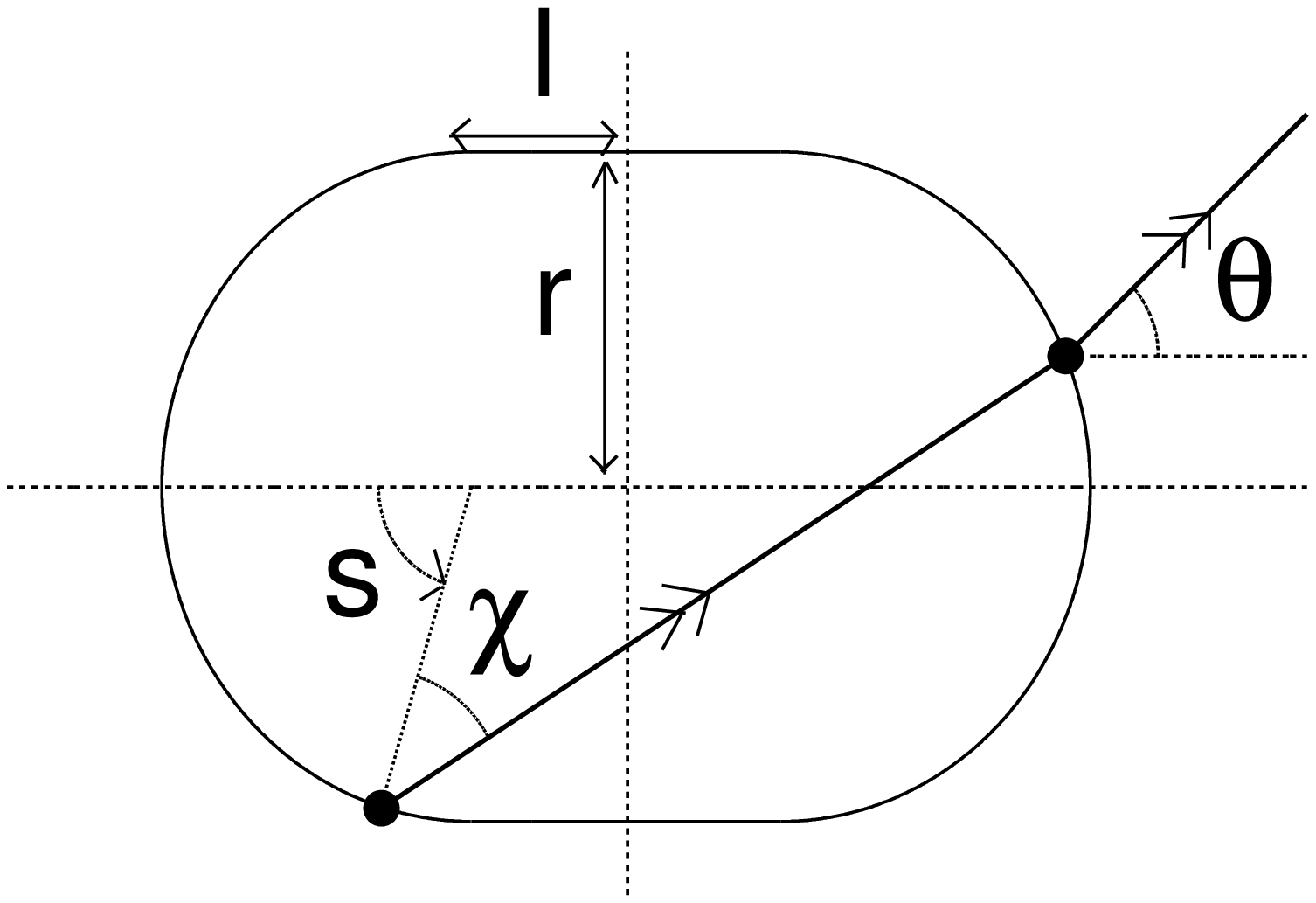}
    \caption{\emph{Left}: Optical microscope image of a stadium-shaped
    micro-laser with $l/r=1$ and  $r~=~30~\mu m$.
    \emph{Right}: Notations. $l$
is the half-length of the rectangle and $r$ is the radius of the
half-circles. $s$ is the curvilinear coordinate along the boundary,
$\chi$ is the incident angle and $\theta$ is the outgoing angle
measured from the main axis of the stadium.}
    \label{dessin}
\end{figure}

Here we focus on polymer micro-lasers of the Bunimovich stadium
shape. Such billiard is the archetype of chaotic systems
\cite{bunimovich}. Stadium billiards look like a rectangle of length
$2l$ between two half-circles of diameter $2r$, accounted the form
ratio $l/r$ (see Fig. \ref{dessin}). Though all periodic orbits are
unstable, dielectric micro-cavities with this shape are well-behaved
lasers with highly directional emission in the far-field pattern
\cite{APL}. Aside from quantum chaos studies, such micro-resonators
are also being considered for applications in integrated optics and
biological sensors \cite{naisbio}.

This letter is devoted to the investigation of the directional
emission versus the form ratio as a way to infer general behavior
for such dynamical systems. First we present experimental results
and show a good agreement with ray numerical simulations. Then we
develop a simple ray model which permits to describe analytically
main observed features. All these results are finally confirmed by
electromagnetic numerical calculations.

\begin{figure}[ht!]
\centering
\includegraphics[width=.8\linewidth]{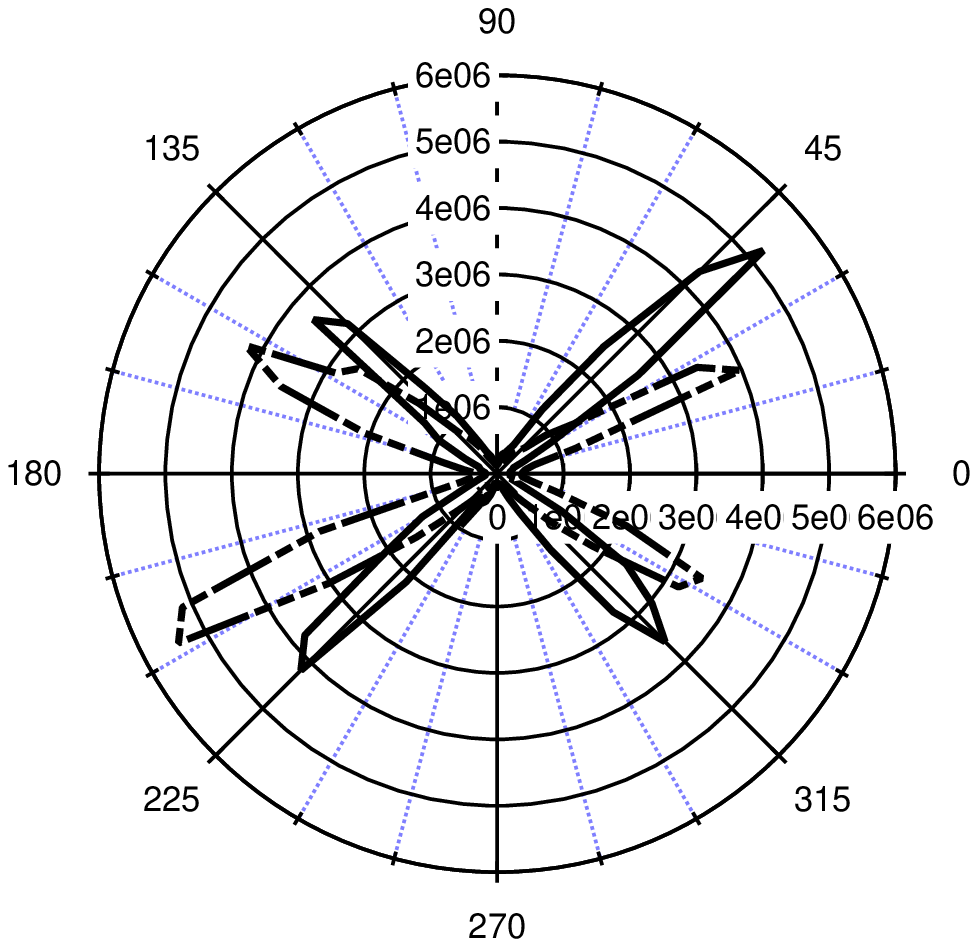}
\includegraphics*[width=.9\linewidth]{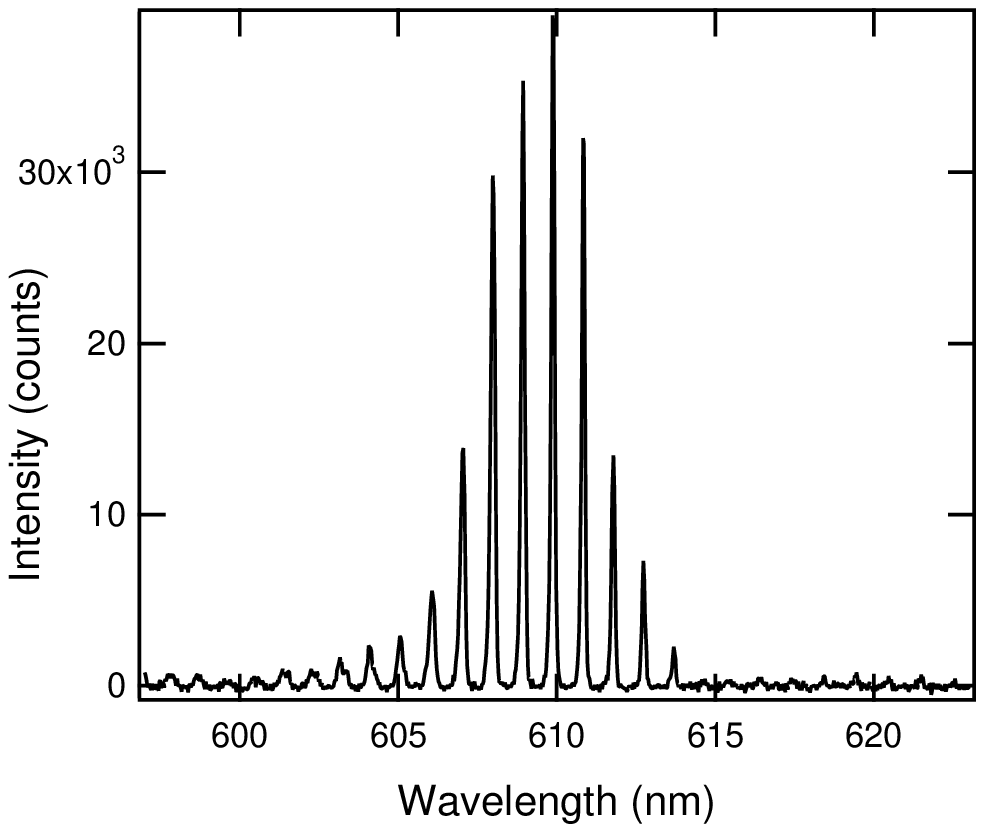}
\caption{\emph{Top}: Experimental far-field emission in the plane of
a 2D stadium-shaped micro-laser as a function of the polar angle for
$l/r=0.5$ (solid line) and $l/r=1$ (dashed line). \emph{Bottom}:
Experimental far-field spectrum for $l/r=2$ and $r=17.5~\mu
m$.}\label{manip}
\end{figure}

The micro-laser cavities are etched in a thin layer of a passive
polymer matrix (PMMA: $polymethyl\-methacrylate$), doped with a
guest laser dye (DCM:
$4-dicyanomethylene-2-methyl-6-(4-di\-me\-thyl\-amino\-styryl)-4H-pyran$).
Our versatile fabrication process ensures for broad variations in
size ($r=10~\mu m$ to $50~\mu m$), shape ($l/r=0$ to $l/r=5$) and
thickness (between $0.4~\mu m$ and $0.7~\mu m$ to avoid vertical
multimode behavior) with quality factor $Q$ greater than 6000
\cite{APL}. The microlasers are pumped uniformly from above at
$532~nm$ with a frequency doubled picosecond Nd:YAG laser and emit
from their sides with $\lambda\simeq 600~nm$. Such 2D dielectric
billiards present an effective refractive index $n$ of 1.5 and are
operating in the semi-classical regime ($r/\lambda$ varying from
typically $20$ up to $100$). The emitted light is collected in the
far-field and coupled to a multi-channel detector via a
spectrometer. A typical experimental spectrum is shown in
Fig.~\ref{manip} (bottom). The sample is fixed on a rotating mount
and the intensity emitted in a given direction is obtained by
summing over all pixels of the spectrum. Far-field spectra are
performed over 360$^{\circ}$ by 5$^{\circ}$ steps with the aperture
angle of about 15$^{\circ}$.

\begin{figure}[ht]
\includegraphics[width=.9\linewidth]{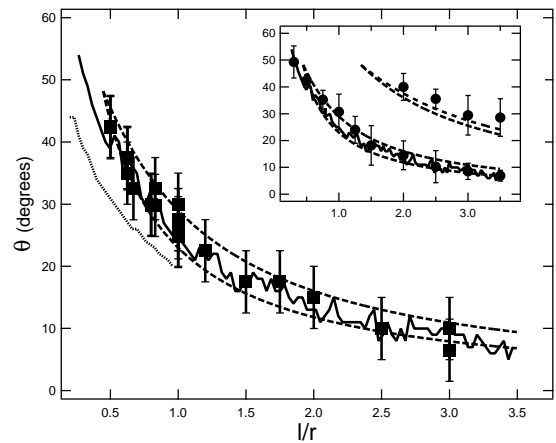}
\caption{Direction of maximal emission $\theta$ versus form ratio
$l/r$. Squares represent experimental results. Solid line indicates
dominant contribution of ray numerical simulations and dotted line
is the position of the largest satellite peak. Dashed lines are lens
model  predictions (\ref{lens}) with $m=0$ (lower line) and $m=-1$
(upper line).
 \emph{Inset:} Comparison of  ray (solid line) and wave numerical simulations (circles).
 Dashed lines are lens model predictions with (from bottom to top)
 $m=0,-1,1,-2$.}
 \label{total}
\end{figure}

The intensity of the laser emission versus the $\theta$ angle (see
Fig.~\ref{dessin} (right) for notations) is displayed in
Fig.~\ref{manip} (top). The emission exhibits a high directionality
along four directions, symmetrical with respect to the 0$^{\circ}$
and 90$^{\circ}$ axis, according to the obvious symmetries of the
stadium shape. The direction of maximal emission depends on the
$l/r$ form ratio. Experimental results are summarized in
Fig.~\ref{total} for form ratios ranging from $l/r=0.5$ to $l/r=3$.
They represent an optimum for reproducibility according to our
5$^{\circ}$ precision interval.

Even if stadium-shaped billiards are fully chaotic systems, these
micro-lasers present experimentally a highly directional emission in
the far-field pattern. The direction of maximal emission depends on
the $l/r$ form ratio and a variation of more than 30$^{\circ}$ has
been measured between $l/r=0.5$ and $l/r=3$ (Fig. \ref{total})what
can be of great interest for applications in integrated optics.

\begin{figure}[ht]
\includegraphics[width=.7\linewidth]{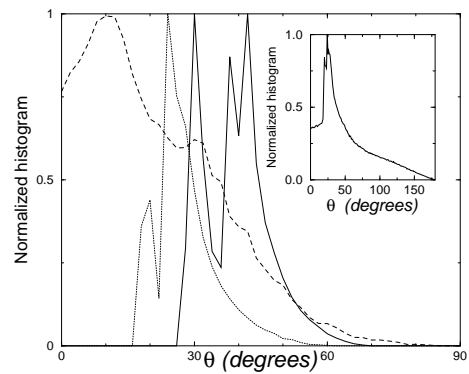}
\caption{Histograms of outgoing angles obtained by ray numerical
simulations with threshold  equal twice the billiard perimeter for
$l/r=0.5$ (solid line), $l/r=1$ (dotted line), and $l/r=2.5$ (dashed
line). Pictures normalized to unit maximum. \emph{Inset}:  The same
for $l/r=0.5$ but without cut off.} \label{histo}
\end{figure}

To account for these experimental results, we first perform usual
ray numerical simulations consisting in probing a large number of
randomly defined rays ($\sim 10^7$) with starting points and initial
directions uniformly distributed over the whole phase space. Each
ray propagates along a straight line and is totally reflected at the
boundary until the incident angle $\chi$ becomes smaller than the
critical angle, $\chi_c=\mathrm{arcsin}(1/n)$, thus allowing the ray
to escape by refraction \footnote{The final result remains
essentially unchanged if a part of the ray, proportional to the
modulus squared of the Fresnel reflection coefficient, is  reflected
into the cavity and keeps propagating.}. The stadium-shaped billiard
is fully chaotic, so almost each ray escapes after propagating over
a finite distance. Typical histograms of the outgoing angles are
plotted in Fig. \ref{histo}.
 To remove the background, only those rays which propagate over more
 than a finite threshold distance (twice the billiard perimeter in
 Fig.~\ref{histo}) are taken into
account. Histograms are indented by peaks which reveal the structure
of underlying unstable manifolds \cite{josab,prlharayama}. An
histogram is calculated for each form ratio value and the position
of its maximum is plotted versus $l/r$ in Fig.~\ref{total} (solid
lines).  The plot is in excellent agreement with real experimental
results as seen from Fig.~\ref{total}.

For some $l/r$ values ray simulations reveal the existence of a few
peaks with comparable amplitudes (cf. Fig.~\ref{histo}). Dotted line
branch in Fig.~\ref{total} indicates the most prominent satellite
peak. Nevertheless dominant experiment contribution is assumed to
correspond to the upper branch because its associated intensity
(integral below the peak) is larger than the other one (cf. solid
line curve in Fig.~\ref{histo}).\\

To explain main features of experiments and ray simulations, we
develop a simple model based on geometrical optics. To minimize
refraction loss, rays propagating near the boundary are privileged.
So we focus on rays which, after being totally reflected on the left
half-circle, come to bounce on the right one (see Fig.~\ref{dessin}).
Long-lived trajectories can be characterized by their
density on the Poincar\'e surface of section with coordinates  $s$
and $\sin \chi$.  Our main assumption is  that this density is
uniform in the allowed part of phase-space
\begin{equation}
P(s,\sin \chi)  \sim \Theta(\frac{\pi}{2} r-|s|)\Theta(|\sin
\chi|-1/n) \label{density}
\end{equation}
where $\Theta(x)$ is the Heaviside step function.

Long-lived trajectories in open chaotic systems are concentrated
near unstable manifolds of confined periodic orbits
\cite{josab,boulanger}. So the true density is a fractal whose
correct determination requires the knowledge of large number of
periodic orbits.  However, finite experimental resolution and
quantum effects will smooth out small details and as the first
approximation the assumption (\ref{density}) looks quite natural.
Below  we show that it leads to simple analytical formulas in a good
agreement with direct ray simulations.

Let us consider  point sources on the left half-circle emitting light
according to (\ref{density}). Among these, some rays escape by refraction
on the right half-circle and one can deduce the histogram of
outgoing angles. The dependence of the histogram maximum on  $l/r$ reproduces
well experiments and ray simulations, validating the assumptions. As the resulting
curves are closed to lens model predictions (see below) we did not present them.

To further simplify  this approach, we consider the right
half-circle as an ideal lens. In this approximation all rays emitted
from a point source are focused exactly as rays close to the line
connecting the source and the center of the right half-circle (i.e.
spherical aberrations are ignored). So the direction of emission
$\theta$ corresponds mainly to this line and can be calculated
analytically from purely geometrical considerations.  It depends on
the emission angle $\chi$ and on the number $|m|$ of rebounds on
straight boundaries ($m>0$ - resp. $m<0$ - means first bounce on the
high - resp. low - straight boundary):
 \begin{equation*}
 \theta(\chi) =|\arcsin \frac{m}{\sqrt{m^2+(l/r)^2}}+
 \arcsin  \frac{\sin \chi}{2\sqrt{m^2+(l/r)^2}}\;|.
 \label{theta}
 \end{equation*}
 This line exists provided $x_1<\sin \chi<x_2$ where $x_1=1/n$ and
 $x_2=2l/\sqrt{4l^2+(2m+1)^2r^2}$. Formally  the channel with given $m$
 is open when $x_1< x_2$ or $l/r>|2m+1|/2\sqrt{n^2-1}$ but it may exist
 for  smaller $l/r$ as well. For a given $l/r$,  the mean emission angle, $\tilde{\theta}$,
 is deduced by averaging over the above interval
 \begin{equation}
 \tilde{\theta}=\frac{1}{x_2-x_1}
 \int_{x_1}^{x_2}\theta(\chi){\rm d}\sin \chi\;.
 \label{lens}
 \end{equation}
Within the lens model the emission intensity  is proportional to the
angle with which the right half-circle is visible from a source
point. Direct geometrical calculations show that the dominant
contribution for $l/r<3.5$ corresponds to $m=0$ channel. Channel
with $m=-1$ has smaller but comparable intensity. Considering
experimental precision, other channels can be ignored but they are
visible in ray simulations (cf. a small peak around $30^\circ$ in
the histogram for $l/r=2.5$ in Fig.~\ref{histo}). The directional
emission predictions for channels with $m=0$ and $m=-1$ are
indicated by dashed lines in Fig.~\ref{total}.  The excellent
agreement between these curves, experiments and direct ray
simulations confirms that the assumption (\ref{density}) is relevant
to interpret analytically  the emission of the stadium-shaped cavity
in the geometrical optics domain.\\

For completeness we perform also numerical calculations in the wave
domain. In the simplest approximation, the electromagnetic field is
evolving in a passive cavity with a defined polarization. In this
case, the electric field (TM polarization) or its magnetic
counterpart (TE polarization) is represented by a scalar wave
function $\psi$ obeying the two dimensional Helmholtz equation
[$(\Delta+n_{i,e}^{2}k^{2})~\psi =0$ where refractive index $n_i=n$
inside and $n_e=1$ outside  the cavity]  with specific boundary
conditions (see e.g. \cite{naisbio}). To find eigenvalues
$\{k_{m}\}$ and eigenfunctions $\{\psi_{m}\}$ of quasi-stationary
states we use the standard boundary element method \cite{wiersig}.
In this approach  the wave functions inside and outside the cavity
$\psi_{i,e}$ are expressed in terms of single layer potentials
$\mu_{i,e}$:
\begin{equation}
\psi_{i,e}(\vec{r}\,)=\oint {\rm d}
s~\mu_{i,e}(s)~H_{0}^{(1)}(kn_{i,e}|\vec r-\vec r_{s}|)
\label{psiext}
\end{equation}
$s$ is the curvilinear coordinate along the boundary.   Here $H_{0}^{(1)}$ is the
Hankel function of the first kind. The choice of this Hankel
function outside the cavity is dictated by outgoing conditions to
infinity
\begin{equation}
\psi_e(r,\theta)\stackrel{r\to\infty}{\longrightarrow} \frac{{\rm e}^{ {\rm i}
kr}}{\sqrt{kr}}f(\theta)
\label{asymptotics}
\end{equation}
and inside the cavity by stability of numerical algorithm.

\begin{figure}[ht]
\centering
\includegraphics[width=.49\linewidth]{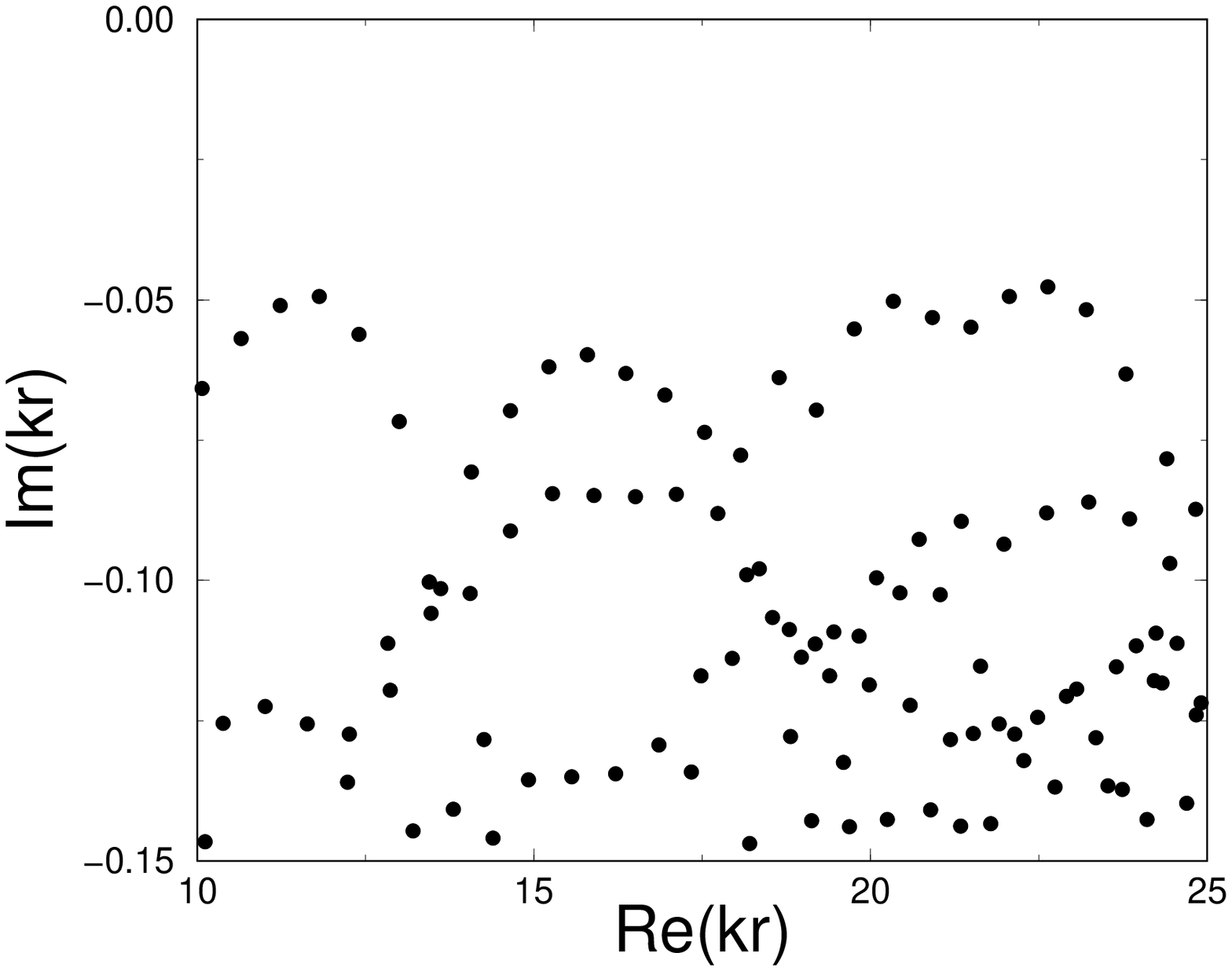}\hfill
\includegraphics[width=.49\linewidth]{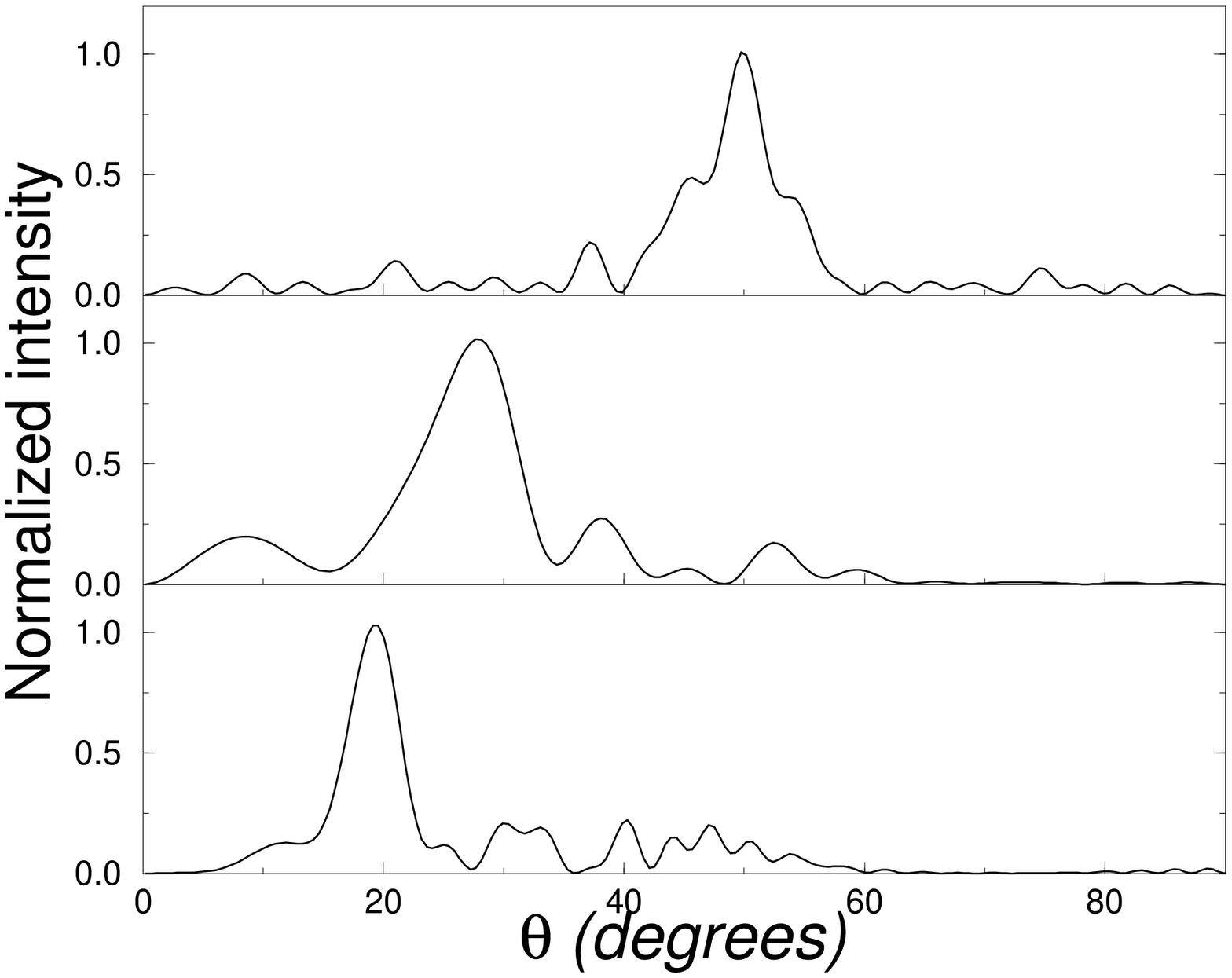}
\caption{\emph{Left}: Imaginary part of  quasi-stationary
eigenvalues versus their real part. These eigenvalues  correspond to
solutions of the Helmholtz equation with TM boundary conditions for
dielectric stadium with  $l/r=2$ and refractive index n=1.5
antisymmetric with respect to both axes. \emph{Right}: Far-field
emission pattern of eigenfunctions: $|f(\theta)|^2$ in
(\ref{asymptotics}) normalized to unit maximum.  From top to bottom
($l/r$=0.3, $Re(kr)$=39.31) ; ($l/r$=1, $Re(kr)$=13.88) ; ($l/r$=2,
$Re(kr)$=23.29).}
    \label{charles}
\end{figure}

Following natural filtering by the lasing effect, only modes with
small losses are considered, i.e. modes with an imaginary part of
the wave-number closed to zero. Most of the well-confined
wave-functions present a clear directional emission in the far-field
pattern, even for low $|kr|$ (see Fig.~\ref{charles} (right)). Then
it is possible to find positions of directional dominant peaks for
each $l/r$ and different eigenvalues which are plotted  in
Fig.~\ref{total} (inset). Each point at fixed $l/r$ corresponds to
the averaging  of maximal emission angles over a representative set
of well-confined eigenfunctions. The error bars corresponds to the
mean standard deviation. As was predicted, dominant emission
direction corresponds to $m=0$ and $m=-1$ channels of the lens
model. It is interesting to notice that channels with $m=1$ and
$m=-2$ are also well visible.

So good agreement between geometrical and wave optics does not
usually appear for closed chaotic systems. The emission directions
of these quantum dielectric billiards behave largely as if they were
classical ones, governed by  geometrical optics rules. However there
is no loss of coherence as in open systems generally studied in the
field of decoherence \cite{decoherence}: experimental spectra
present clear peaks (see fig.~\ref{manip} (bottom)) and no noise is
introduced to couple light with the outside of the cavity.
 Actually the typical distance $L/r$ covered by a photon - of the
order of $2Q/n|k|r$ - ranges between 16 and 80 in our experiments.
This magnitude suggests that interference effects should be taken
into account whereas we have demonstrated that a geometrical optics
approach is sufficient to predict the overall directional emission.
The lens model shows that it is mostly connected with the
existence of only a small number of well defined open emission
channels. Under higher precision calculations and experiments, each
channel should split into many narrow peaks and quantum phenomena
should be more important.

In summary, we  present  experimental results about directional
emission of stadium-shaped micro-lasers and demonstrate that they
are well described by simple geometrical optics approaches.  We
develop a lens model which predicts analytically main features of
directional emission.  Moreover we point out that well-confined
wave-functions show a good agreement with these classical
predictions  while coherent properties are not destroyed.

These kinds of predictions seem to be applicable to a more general
class of chaotic dielectric billiards. Experiments and calculations
are thus under progress to explore other   micro-cavity shapes  like
cut disk and cardioid.

The authors are grateful to R. Hierle, D. Wright and S. Brasselet
for experimental and technological support and to O. Bohigas and T.
Nguyen for helpful discussions.


\begin{thebibliography}{99}

\bibitem{ecoles}
{\it Chaos and Quantum Physics}, Proceeding of the 1989 Les Houches
Summer School, edited by M. J. Giannoni, A. Voros, and J.
Zinn-Justin (Elsevier, 1991); {\it New Directions in Quantum Chaos},
Proceedings of the 1999 Varenna Summer School, edited by G. Casati,
I. Guarneri, and U. Smilansky (IOS Press, 2000).

\bibitem{microondes}
J. Stein, H.-J. St\"ockmann, and U. Stoffregen, {\it Phys. Rev.
Lett.}, {\bf 75}, 53 (1995); H. Alt, C. Dembowski, H.-D. Gr\"af, R.
Hofferbert, H. Rehfeld, A. Richter, and C. Schmit, {\it Phys. Rev.
E}, {\bf 60}, 2851 (1999).

\bibitem{Davidson}
M. F. Andersen, A. Kaplan, and N. Davidson, {\it Phys. Rev. Lett.},
{\bf 90}, 023001 (2003).

\bibitem{decoherence}
 S. Habib, K. Shizume, and W. H. Zurek, {\it Phys. Rev.
Lett.}, {\bf 80}, 4361, (1998).

\bibitem{cvitanovic}
 P. Cvitanovic and B. Eckhardt, {\it Phys. Rev. Lett.}, {\bf 63},
823 (1989);  K. Pance, W. Lu, and S. Sridhar, {\it Phys. Rev.
Lett.}, {\bf 85}, (13), 2737 (2000).

\bibitem{boulanger}
J.P. Keating, M. Novaes, S.D. Prado, and M. Sieber,
quant-ph/0605217; S. Nonnenmacher and M. Rubin, in preparation.

\bibitem{random}
X. Jiang, S. Feng, C. M. Soukoulis, J. Zi, J. D. Joannopoulos, and
H. Cao, {\it Phys. Rev. B}, {\bf 69}, 104204 , (2004).

\bibitem{copains}
C. Gmachl, F. Capasso, E. E. Narimanov, J. U. N\"ockel, A. D. Stone,
J. Faist, D. L. Sivco, and A. Y. Cho, {\it Science}, {\bf 280}, 1556
(1998); T. Harayama, P. Davis, and K. S. Ikeda, {\it Phys. Rev.
Lett.}, {\bf 90}, 063901, (2003); A. D. Stone, {\it Physics Today},
{\bf 58}, (8), 37 (2005).

\bibitem{bunimovich}
L. A. Bunimovich, {\it Commun. Math. Phys.}, {\bf 65}, 295
(1979).

\bibitem{APL}
 M. Lebental, J.-S. Lauret, R. Hierle, and J. Zyss, {\it Appl.
Phys. Lett.}, {\bf 88}, 031108 (2006).

\bibitem{naisbio}
Proceedings of the International School of Quantum Electronics,
edited by F. Michelotti, A. Driessen, M. Bertolotti, {\it AIP
conference proceedings}, {\bf 709} (2004).

\bibitem{josab}
H. G. L. Schwefel, N. B. Rex, H. E. Tureci, R. K. Chang, A. D.
Stone, T. Ben-Messaoud and J. Zyss, {\it Jour. Opt. Soc. Am. B },
{\bf 21}, 923 (2004).

\bibitem{prlharayama}
S. Shinohara, T. Harayama, H. E. Tureci and A. D. Stone,
physics/0606212.

\bibitem{wiersig}
J. Wiersig, {\it Phys. Rev. A}, {\bf 67}, 023807 (2003).

\end{thebibliography}
\end{document}